\documentclass[12pt]{article}

 \usepackage{amssymb}
 \usepackage{amsmath}
 \usepackage{amsfonts}
 \usepackage{color}
 \newcommand {\be}{\begin{equation}}
 \newcommand {\ee}{\end{equation}}
 \newcommand {\bea}{\begin{array}}
 
 \newcommand {\eea}{\end{array}}

 \newcommand {\bx}{\textbf{x}}
\newcommand {\bk}{\textbf{k}}

\begin{document}

  \thispagestyle{empty}

  \vspace{2cm}

  \begin{center}
    \font\titlerm=cmr10 scaled\magstep4
    \font\titlei=cmmi10 scaled\magstep4
    \font\titleis=cmmi7 scaled\magstep4
  {\bf Quantization In Finite Volumes \\
      Using Symplectic Quantization Programm}

    \vspace{1.5cm}
    \noindent{{
        Sh. Chenarani\footnote{s.chenarani@ph.iut.ac.ir}
        A. Shirzad\footnote{shirzad@ipm.ir}
         }}\\
    \vspace{0.8cm}

   {\it Department of Physics, Isfahan University of Technology \\
P.O.Box 84156-83111, Isfahan, IRAN, \\
School of Physics, Institute for Research in Fundamental Sciences (IPM)\\
P.O.Box 19395-5531, Tehran, IRAN.}

  \end{center}

  \vskip 2em

\begin{abstract}
We use the ideas of symplectic quantization for quantizing fields in
finite volumes. We consider, as examples, the Klein-Gordon and
electromagnetic fields in three different boxes. As a second idea we
consider the given boundary conditions as primary constrains.
Consistency of primary constrains leads to infinite chains of
constraints at the boundaries. Without solving the equation of
motion, we impose the set of constraints on suitable expansions of
the fields. We show that if the new set of variables, such as
Fourier modes, are chosen appropriately, imposing the constraints
omits a few number of canonical pairs. Hence, the reduced phase
space, with canonical pairs as coordinates, is achieved.
\end{abstract}



\section{Introduction \label{sec1}}
Every student of quantum field theory may have encountered the
question "what is the reason the people consider the coefficients of
Fourier expansions of the fields as creation and annihilation
operators and why this assumption works?". A possible answer is:
"one can retain the original canonical brackets among the fields by
using the assumed creation-annihilation algebra".

Such a statement does not show the essential meaning inherent in the
quantum modes and the way the assumed algebra has been emerged.
Moreover, it should be explained how the constraint equations as
well as the gauge conditions are remained valid as relations among
the operators. There exist also difficulties with boundary
conditions. The problem is the canonical algebra among the fields
are violated on the boundaries. \footnote{As an example, the
contradiction between the canonical Poisson brackets and the mixed
boundary conditions in the model of bosonic string in the background
B-field leaded people to new ideas such as non-commutative
coordinates of the string at the boundaries\cite{ref4,ref0} } The
reason is the boundary conditions and their consistency in time
constitute a set of second class constraints which change the
Poisson brackets to Dirac brackets.

Hence we come to conclusion that the canonical quantization programm
should be managed in a more convincing way within the framework of
Dirac constraint theory\cite{Dirac,batelle,hennax}. This programm is
a systematic and universal quantization method based on the
canonical structure of the classical system. This framework is
inevitably necessary for singular Lagrangians such as gauge systems.
The boundary conditions viewed as essential identities among the
coordinates and (velocities or) momenta should also be considered as
Dirac constraints where their consistency in time may give more
constraints.

The whole procedure leads us to omitting redundant variables
considered initially in the model due to the symmetry structure of
the theory. Hence, one finds the "pure" space of physical variables
which is called the "reduced phase space". These variables are the
only meaningful things to be considered as quantum operators. This
procedure is often escaped in the textbooks on quantum field theory.
Hence, some physical concepts may be sacrificed or remain unclear in
such "operational methods", since they use "ad-hoc quantization
assumptions" emphasizing only on the results.

One approach to canonical quantization is based on Fadeev-Jackiev
\cite{FJackiw} method of analyzing constrained systems and is
recognized in the related literature \cite{mojiri} as "symplectic
quantization". In ref. \cite{bakhshi} the basic concepts and proofs
of the "symplectic quantization programm" are proposed and applied
to the model of a massive bosonic string in a background B-field.
The essential aspects of the symplectic quantization can be
summarized in the following steps:

1) Investigating the constraint structure of the system, including
the boundary conditions and their consistency in time;

2) fixing the gauges by imposing appropriate gauge fixing
conditions;

3) choosing suitable coordinates for the phase space to impose the
constraints, finding the physical modes as the coordinates of the
reduced phase space and writing the expansions of the fields and
momentum fields in terms of them;

4) inserting the above expansions in the symplectic two-form;

5) inverting the symplectic matrix of the reduced phase space to
find the Dirac brackets of the physical modes and converting them
to quantum commutators;

6) writing the Hamiltonian in terms of the physical modes, finding
their equations of motion and then writing the physical modes in
terms of their initial values (Schrodinger modes);

7) writing the expansions of the fields in terms of Schrodinger
modes, and finding their time dependent commutation relations by
using the commutator algebra of Schrodinger modes.

According to symplectic approach (steps 4 and 5) for a theory with
dynamical fields $\phi_s(x,t)$ and momentum fields $\pi_s(x,t)$ in
$d+1$ dimensions the symplectic two-form is defined as
 \begin{equation} \Omega=2\sum_s\int d^dx
d\pi_s(x,t)\wedge d\phi_s(x,t). \label{w-14}
 \end{equation}
Under imposing the constraints (and gauge fixing conditions) the
original fields can be written in terms of physical modes $a^n(t)$
which are the coordinates of the reduced phase space. These are the
main physical quantities of the theory. Inserting the expansions of
the original fields and momentum fields in the symplectic two-form,
it can be written as
\begin{equation}
\Omega=\sum_{n,m}\omega_{nm}da^n\wedge da^m. \label{w-15}
\end{equation}
It has been shown \cite{bakhshi} that the antisymmetric tensor
$\omega^{mn}$ which is the inverse of the symplectic tensor
$\omega_{nm}$ defines the classical brackets (in fact the Dirac
brackets) among the physical modes as
\begin{equation}
\{a^n,a^m\}=\omega^{nm} \ .\label{w-16}
\end{equation}

In this paper we want to show that "the symplectic quantization
programm" gives the desired answer for some familiar field theories
in finite volumes. Our emphasize is on the "method" and not on the
"results" (which we expect anyhow to be consistent with the
literature). Our main goal is to show that the standard quantization
idea of Dirac (converting the Dirac brackets to commutators) is
enough for each model and it is not needed to "assume" alternative
"quantization assumptions" in different cases. We use this program
for the following models: Klein-Gordon field in a rectangular box, a
long cylinder, and a sphere; and the electromagnetic field inside a
rectangular box, a cylindrical box and a sphere. In the following
section we explain our method in more details by investigating the
klein-Gordon field in a rectangular box. Then other examples are
considered more briefly in the subsequent sections. In the last
section we will give our concluding remarks.

\section{Klein-Gordon field in rectangular box}
The classical action of the Klein-Gordon field is
\be S=\frac{1}{2}\int \left[\partial_{\mu}\phi\partial^\mu
\phi-m^2 \phi^2 \right]d^4x\ . \label{a-1} \ee
Suppose our system is confined inside a rectangular box with
boundaries at $ x=0\  ,a\ \  ; y=0\  ,b\ \ ; z=0\  ,c$. Using the
Lagrangian
 \be
 L=\frac{1}{2}\int{d}^3{x}\left[\dot\varphi^2-\nabla{\varphi}^2
 -m^2\varphi^2)\right],
 \label{a-3} \ee
the momentum field is $\pi(\bx)=\dot{\varphi}(\bx)$ and the
canonical Hamiltonian reads
 \be
 H_c{=}\frac{1}{2}\int{d}^3{x}\left[\pi^2+\nabla{\varphi}^2
 +m^2\varphi^2)\right].\label{a-4}
 \ee
Suppose we are given Dirichlet boundary condition on the
boundaries which can be considered as the following primary
constraints
\be \begin{array}{lll}\varphi(0,y,z)\approx 0, & \varphi(x,0,z)\approx 0,
&  \varphi(x,y,0)\approx 0, \\
\varphi(a,y,z)\approx 0,  &   \varphi(x,b,z)\approx 0, &
\varphi(x,y,c)\approx 0 ;
\end{array} \label{a-5}
 \ee
where the symbol $\approx$  means weak equality. The total
Hamiltonian is symbolically
 \be
H_T=H_c+\sum_{i=1}^{6}\int\lambda_i\psi_i^{(0)},
 \label{a-6}
 \ee
where $\psi_i^{(0)} $    are the six constraints given in
Eqs.(\ref{a-5}), and the integration is over the remaining
coordinates on each boundary. For example the $i=1$ term is $\int dy
\, dz \lambda_1(y,z)\varphi(0,y,z) $, etc.

As in any constrained systems, we should investigate the consistency
of primary constraints in the course of the time. Consider, for
instance, the consistency  of the constraint
$\psi_1^{(0)}(y,z)=\varphi(0,y,z)$ which using Eq.(\ref{a-6}) and
the fundamental Poisson brackets,
 \be\begin{array}{l}
 \{\varphi(\bx),\pi(\bx')\}=\delta^3(\bx-\bx'),\\
 \{\varphi(\bx),\varphi(\bx')\}=0 ,\\
 \{\pi(\bx),\pi(\bx')\}=0,\end{array} \label{a-7} \ee
gives
 \be \chi_1^{(0)}(y,z)=\{\psi_1^{(0)}(y,z),H_T\}
 =\pi(0,y,z)\approx 0.  \label{a-8}
 \ee
In this way we find second level constraints $\chi_1^{(0)}, \cdots
\chi_6^{(0)}$ for momentum field $\pi(x,y,z)$ at the boundaries,
similar to $\varphi_1^{(0)}, \cdots \varphi_6^{(0)}$ in
Eqs.(\ref{a-5}). At this step we have found 6 constraints
$\chi_1^{(0)}, \cdots \chi_6^{(0)}$ that are conjugate to 6 primary
constraints $\psi_1^{(0)}, \cdots \psi_6^{(0)}$. In the context of
the theory of constrained system, the process of consistency of
constraints should end here since the Lagrange multipliers would be
determined by consistency of secondary constraints  $\chi_i^{(0)}$.
However, the situation is different for the constraints emerged from
boundary conditions. Here, the consistency of $\chi_1^{(0)}(y,z)$ ,
for instance, gives
 \be
 \dot\chi_1^{(0)}(y,z)=\{\pi(0,y,z),H_C\}+\int{dy'dz'}\lambda_1(y',z')
 \{\pi(0,y,z),\varphi(0,y',z')\}\approx0\label{a-9}
 \ee
The first  term at the right hand side of Eq.(\ref{a-9}), after
straightforward calculations using Eqs.(\ref{a-4}) and (\ref{a-7}),
is equal to $\nabla^2\phi(x,y,z)|_{x=0}$. The second term, however,
needs a little care, due to Poisson brackets at the sharp boundary
on $x=0$. To this end, we use delta functions to convert the fields
within the bracket to their usual form. In this way the
corresponding term reads
 \begin{eqnarray}\int dx dx'dy'dz'\lambda_1(y',z')\delta(x)\delta(x')
\{\pi(x,y,z),\varphi(x',y',z')\}=\hspace{3cm}&& \nonumber\\ -\int dx
d^3 x'\lambda_1(y',z')\delta(x)\delta(x')\delta^3(x-x') &&
 \label{a-10} \end{eqnarray}
If we consider Eq.(\ref{a-9}) as an equation to determine the
Language multiplier $\lambda_1(y,z)$, then Eq.(\ref{a-10}) shows
that the coefficient of the unknown is one order more singular than
the first term in Eq.(\ref{a-9}), (see more details in \cite{ref4}
by regularizing the delta functions in Eq.(\ref{a-10})). The only
way to satisfy the consistency condition (\ref{a-9}) is
$\lambda_1(y,z)$ vanish for all $(y,z)$ and at the same time the
first term in Eq.(\ref{a-9}) is considered as a new constraint.
Hence, we find the third level constraints ${\psi_1}^{(1)}, \cdots
{\psi_6}^{(1)}$ as $\nabla^2{\psi_1}, \cdots \nabla^2{\psi_6}$ at
the corresponding boundaries while the lagrange multiplier
$\lambda_i$ has been determined to be zero. In this way the
consistency procedure is not terminated, although the lagrange
multipliers are determined. Next we should impose the consistency of
third level constraints ${\psi_1}^{(1)}, \cdots {\psi_6}^{(1)}$.
Direct calculation using Eqs.(\ref{a-4}) and (\ref{a-7}) shows that
the forth level constraints are $\chi_1^{(1)}, \cdots \chi_6^{(1)}
\equiv \nabla^2 \pi |_{\textrm{boundaries}}$.

The consistency procedure continues  unlimitedly leading to the
following infinite chains of constraints at the boundaries.
\be \begin{array}{l}{\psi_i}^{(n)}=(\nabla^2)^n \varphi
|_{\textrm{boundaries}} \\ {\chi_i}^{(n)}=(\nabla^2)^n \pi
|_{\textrm{boundaries}}\label{a-11} \end{array} \ee
where $n$ is any integer and the index i refers to six boundaries
involved.

Next we want to see how the constraints (\ref{a-11}) restricts the
physical degrees of freedom. In its original form, the phase space
of the theory consists of the field variable $\varphi(x,y,z)$  and
$\pi(x,y,z)$ at all points within the corresponding cubic box, up to
the constraints given in Eqs.(\ref{a-11}). As is well-known, the
Dirac procedure of quantization of such a theory requires
transforming the Dirac brackets of the fields to quantum
commutators. This procedure involves inverting the matrix of second
class constraints. A little look at Eqs. (\ref{a-11}) shows that for
our case we have a system of second class constraints with a
complicated infinite  matrix of Poisson brackets, whose elements are
derivatives of different orders of delta functions at the
boundaries. It is practically impossible to compute such a matrix
and invert it. \footnote{ In refs. \cite{ref4} and \cite{ref0} the
authors have tried to do this for a bosonic string in a background
B-field. But it seems practically impossible to find the answer
without using the results known from other approaches.} However, it
is possible to quantize the theory in a simpler way by changing the
dynamical variables.

As explained in details in \cite{ref3} , in
our case for instance, one prefers to do a canonical
transformation from the original variables $\phi (\textbf{x})$ and
$\pi (\textbf{x}) $ to a new set of suitable canonical coordinates
in which the constraints (\ref{a-11}) lead to omitting a number of
canonical coordinate-momentum pairs. This should finally leave us
with a reduced phase space in which the remaining canonical pairs
act as the physical degrees of freedom. The Fourier transformation
sometimes dose this job for models with Dirichlet or Neuman
boundary conditions. In fact, this is the mysterious behind the
fact  that Fourier transformation is, most of the time, the first
step toward the aim of quantization of the fields.

Let us consider the following Fourier transformations of the
original fields.
\be \varphi(\bx,t)=\frac{1}{(2\pi)^{3/2}}\int a(\bk,t) e^{ikx}d^3
k ,\label{a-12} \ee
\be\pi(\bx,t)=\frac{1}{(2\pi)^{3/2}}\int b(\bk,t) e^{-ikx}d^3 k.
\label{aa-12} \ee
Using the inverse Fourier transformations and the fundamental
Poisson brackets (\ref{a-7}) we can find the Poisson brackets of
the new variables $a(\bk,t)$ and $b(\bk,t)$ as

\be\begin{array}{l}
 \{a(\bk,t),b(\bk',t)\}=\delta^3(\bk-\bk'),\\
 \{a(\bk,t),a(\bk',t)\}= 0 , \\ \{b(\bk,t),b(\bk',t)\}= 0.
 \label{a-13}
 \end{array} \ee
This shows that the Fourier transformation given in Eqs.
(\ref{a-12}) and (\ref{aa-12}) is a canonical transformation. Now
let us impose the constraints (\ref{a-11}) on the expansions
(\ref{a-12}) and (\ref{aa-12}) to omit nonphysical modes. Imposing,
for instance, the constraints ${\psi_1}^{(n)}=(\nabla^2)^n \varphi
|_{x=0}$ on $\varphi(\bx,t)$ gives
\be \int a(\bk,t) (k^2)^n e^{i(k_y y+k_z z)}d^3 k =0. \label{a-14}
\ee
This condition can be satisfied for arbitrary integer $n$ and all
points $(y,z)$ if $a(k_x,k_y,k_z,t)$ is an odd function of $k_x$.
Imposing this restriction on Eq.(\ref{a-12}) shows that the term
$\cos k_x x$  in $e^{ik_x x}$ should be absent. The same thing
happens for $\pi(\bx,t)$ and also for the terms $\cos k_y y$ and
$\cos k_z z$. We are left finally just with the $\sin$ terms in
the expansions (\ref{a-12}) and (\ref{aa-12}) as
\begin{eqnarray}&&\varphi(\bx,t)=\frac{1}{(2\pi)^{3/2}}\int
a(\bk,t) \sin k_x x\sin k_y y \sin k_z z d^3 k \label{a-15} \\&&
\pi(\bx,t)= \frac{1}{(2\pi)^{3/2}}\int b(\bk,t) \sin k_x x\sin k_y
y \sin k_z z d^3 k\label{aa-15}\end{eqnarray}
where $a(\bk,t)$ and $b(\bk,t)$ are odd functions of $k_x$, $k_y$
and $k_z$. Imposing the constraints ${\psi_4}^{(n)} \equiv
(\nabla^2)^n \varphi(x)|_{x=a}$ on the expansion of
$\varphi(\bx,t)$ in Eq.(\ref{a-12}) gives
\be \int a(\bk,t) k^{2n}\sin k_x a\sin k_y y \sin k_z z d^3 k =0
.\ee
This equation should be satisfied for arbitrary integer $n$ and
for all points $(y,z)$. Since the integrand is even with respect
to $k_x$, the only possibility is that $a(\bk,t)=0$ when $\sin k_x
a\neq0$. In other words, we are left with modes of the form
$k_x=p\pi x/a$, where $p$ is some integer. The same argument for
boundary conditions on the surfaces $y=b$ and $z=c$, and for the
momentum field as well, gives the following expansion for the
fields;
\begin{eqnarray}&& \varphi(\bx,t)=\sum_{p,r,s} A_{prs}(t)\sin \frac{p\pi x}{a}
\sin\frac{r\pi y}{b}\sin\frac{s\pi z}{c}\ ,\label{a-16} \\ &&
\pi(\bx,t)=\sum_{p,r,s} B_{prs}(t)\sin\frac{p\pi
x}{a}\sin\frac{r\pi y}{b}\sin\frac{s\pi z}{c}\ . \label{aa-16}
\end{eqnarray}
Since we were left previously with odd functions $a(\bk,t)=0$ and
$b(\bk,t)=0$, the summations in Eqs.(\ref{a-16}) and (\ref{aa-16})
are only on positive integers.

As we see the reduced phase space is much smaller than the
original one. We began with the fields $\varphi(\bx,t)$ and
$\pi(\bx,t)$  which introduce innumerably infinite number of
degrees of freedom. Going to the Fourier modes $a(\bk,t)$ and
$b(\bk,t)$, half of  degrees of freedom are killed because of the
constraints at $x=0$, $y=0$  and $z=0$ surfaces, and within the
remaining ones most of them are omitted due to the constraints at
$x=a$, $y=b$ and $z=c$ surfaces. Finally we are left with
infinitely numerable degrees of freedom labeled by the positive
integers $(p,r,s)$.

We emphasize that in order to recognize the space of physical
variables we need not to solve the full dynamics of the system. In
other words, we have considered so far just the dynamics of the
constraints. If we force the fields in Eqs.(\ref{a-16}) and
(\ref{aa-16}) to satisfy the equations of motion, then $A_{prs}(t)$
and $B_{prs}(t)$ should have definite time dependence. Upon
quantization we can find the commutation relations among the
physical degrees of freedom $A_{prs}(t)$ and $B_{prs}(t)$, while
their evolution during time depends on the specific from of the
Hamiltonian.

Now let us find the brackets of the physical modes $A_{prs}(t)$ and
$B_{prs}(t)$ using the symplectic approach. The symplectic two-form
$ \Omega =\int d^3 x d \pi(\bx,t)\wedge d \varphi(\bx,t)$ in terms
of the physical modes $A_{prs}(t)$ and $B_{prs}(t)$ is
\be \Omega=\sum_{prs} \frac{abc}{4}d B_{prs}\wedge d A_{prs}\ .
\ee
Comparing this with Eq. \eqref{w-15} shows that $\omega= (abc/8)J$
where
\be J=\left( \begin{array}{cc} \textbf{0}&\textbf{-1}
\\\textbf{1}&\textbf{0} \end{array} \right)\ . \ee
$J$ is the standard symplectic matrix with entries written in $AA$,
$AB$, $BA$ and $BB$ blocks respectively. Since $J^{-1}=-J$ the
(Dirac) brackets of physical modes are as follows;
\be
\left\{A_{prs}(t),B_{p'r's'}(t)\right\}=\frac{8}{abc}\delta_{pp'}
\delta_{rr'} \delta_{ss'}. \label{b-20} \ee
Using the above brackets and the final expansions in Eqs.
(\ref{a-16}) the equal time brackets of the original fields are
\begin{eqnarray}&& \left\{\varphi(\bx,t),\pi(\bx',t)\right\}_{\textrm{DB}}=
\frac{8}{abc}\sum_{prs}\sin \frac{p\pi x}{a}\sin\frac{p\pi
x'}{a}\sin\frac{r\pi y}{b} \sin\frac{r\pi y'}{b}\sin\frac{s\pi z}{c}
\sin\frac{s\pi z'}{c}\nonumber \\&& \left\{ \varphi(\bx,t),
\varphi(\bx',t) \right\}_{\textrm{DB}}=0  \label{b-18} \\&&
\left\{\pi(\bx,t),\pi(\bx',t)\right\}_{\textrm{DB}}=0 \nonumber
\end{eqnarray}
It is easily seen that the non vanishing Dirac bracket can be
written as follows
\be \left\{\varphi(\bx,t),\pi(\bx',t)\right\}_{\textrm{DB}}= \left\{
\begin{array}{ll}  \delta^3(\bx-\bx') \hspace{1cm} &
\textrm{inside the box} \\ 0& \textrm{on the boundaries.}
\end{array}\right. \label{bb-19} \ee
It is important to note that the boundary conditions have changed
the original Poisson brackets given in Eqs. (\ref{a-7}) to Dirac
brackets of Eqs. (\ref{bb-19}). Now everything is ready to take our
final step and quantize the system. {\it Upon quantization the
physical modes $A_{prs}(t)$ and $B_{prs}(t)$ should be considered as
quantum operators and the Eqs. \eqref{b-20}-\eqref{bb-19} with right
hand sides multiplied by $(i\hbar)$ as commutation relations.}

We can compute the canonical Hamilton in the reduced phase space
by inserting expansions (\ref{a-16}) in Eq. (\ref{a-4}). The
answer is
\be H_c=\frac{abc}{16}\sum_{prs}\left[{B_{prs}}^2+ \omega_{prs}^2
{A_{prs}}^2 \right],\label{b-1} \ee
where
\be \omega_{prs}^2=m^2+\left(\frac{p^2 \pi^2}{a^2}+\frac{r^2
\pi^2}{b^2}+\frac{s^2 \pi^2}{c^2}\right). \ee
Eq.(\ref{b-20}) shows that the Hamiltonian (\ref{b-1}) is a
superposition of infinite number of simple harmonic oscillators.
Hence, the equations of motion for the physical modes read: $\dot
A_{prs}=B_{prs}$ and $\dot B_{prs}=-\omega^2_{prs}A_{prs}$, which
acquire the following solutions;
\begin{eqnarray} && A_{prs}(t)=A_{prs}(0)\cos (\omega_{prs}
t) +\frac{B_{prs}(0)}{\omega_{prs}}\sin (\omega_{prs} t) \label{cc-1} \\
&& B_{prs}(t)=-A_{prs}(0)\omega_{prs}\sin (\omega_{prs} t) +
B_{prs}(0) \cos (\omega_{prs} t) \label{cc-2}\end{eqnarray}
Inserting Eqs.\eqref{cc-1} and \eqref{cc-2} into Eqs.\eqref{a-16}
and \eqref{aa-16} gives original fields in terms of the Schrodinger
modes $A_{prs}(0)$ and $B_{prs}(0)$. Since the algebra of the modes
given in Eq.\eqref{b-20} is independent of time \cite{bakhshi} the
Poisson brackets of the Schrodinger modes is the same as time
dependent modes. Note that the Schrodinger modes are in fact the
coefficients of the expansions of the fields in terms of the
solutions of the equations of motion, i.e. the Klein-Gordon equation
\be \frac{\partial^2\phi}{\partial t^2}-\nabla^2\phi+m^2\phi=0 .
\label{aa1} \ee
As explained in the introduction, the quantization may be achieved
by "assuming" suitable algebra among the Schrodinger modes (as is
almost done in the literature on quantum field theory). However,
instead of an ad-hoc assumption which works well, we have derived
the algebra of Schrodinger modes in a systematic way. Note also that
the Klein-Gordon field $\varphi(\bx,t)$ is non commutative at
different points of space, as expected.
 \section{Klein-Gordon field inside a cylinder}
Consider the Klein-Gordon field inside an infinite cylinder of
radius $a$, along the z-axis. The Lagrangian and canonical
Hamiltonian is as before (Eqs.\ref{a-3} and \ref{a-4}), while the
primary constraint is given by $\phi(a,\varphi,z)\approx 0$.
Consistency of primary constraint, using the total Hamiltonian
\be H_T=H_c+\int d \varphi dz \lambda (\varphi,z)
\phi(a,\varphi,z) \label{a-19} \ee
and the fundamental Poisson brackets in cylindrical coordinates
gives the second level constraint $\pi (a,\varphi,z)$. Consistency
of $\pi (a,\varphi,z)$ results to vanishing of $\lambda (\varphi,z)$
and at the same time emerging the third level constraint $\nabla ^2
\phi \mid _{\rho =a}=0 $. This happens due to the same reason as
stated after Eqs.(\ref{a-10}).  Fourth level constraint is also
derived as $\nabla ^2 \pi \mid _{\rho =a}=0 $. The process continues
to give two sets of constraints $(\nabla ^2 )^n \phi \mid _{\rho
=a}=0 $ and $(\nabla ^2 )^n \pi \mid _{\rho =a}=0 $.

Let us expand our canonical fields, in the most general form, as the
following Bessel-Fourier integrals;
\begin{eqnarray} &&
\phi(\rho,\varphi,z,t)=\sum_{m=-\infty}^\infty \int_{-\infty}^\infty
d \lambda \int_{-\infty}^\infty d k A_m (\lambda,k,t) e^{i \lambda
z} e^{im\varphi} J_m (k \rho),\label{aa-20} \\&&
\pi(\rho,\varphi,z,t)=\sum_{m=-\infty}^\infty \int_{-\infty}^\infty
d \lambda \int_{-\infty}^\infty d k B_m (\lambda,k,t) e^{-i \lambda
z} e^{-im\varphi} J_m (k \rho). \label{a-20}
\end{eqnarray}
These expansions can be considered as a suitable canonical
transformation from the original variables $\phi (\rho ,\varphi
,z)$ and $\pi (\rho ,\varphi ,z)$ to $A_m (\lambda ,k)$ and $B_m
(\lambda ,k)$ which are more compatible with the symmetry of the
system.

Now imposing the set of constraints ${\nabla}^{2j} \phi|_{\rho=a}
=0$ on $\phi(\rho ,\phi ,z)$ gives
\be \sum_m \int d\lambda d k A_m(\lambda,k,t) e^{i \lambda z} e^{im
\varphi} (-k^2-\lambda^2)^j J_m (k \rho)|_{\rho=a}  =0\ ,
\label{a-23} \ee
by using the Bessel equation, $(\frac{d^2}{d\rho^2} +\frac{1}{\rho}
\frac{\partial}{\partial \rho}-\frac{m^2}{\rho^2})J_m (k
\rho)=-k^2J_m (k \rho)\ $. This equation for arbitrary $j$ shows
that $A_m(\lambda ,k,t)$ should vanishes except for values of $k$
that
\be J_m (k a)=0 \Rightarrow k=k_{mn}\equiv \frac{x_{mn}}{a}, \ee
where $x_{mn}$ are the roots of the Bessel function $J_m(x)$. The
same thing happens for $\pi(\rho ,\varphi ,z,t)$ and we can write
the following expansion for the fields;
\begin{eqnarray} &&
\phi(\rho,\varphi,z,t)=\sum_{mn} \int d \lambda  A_{mn}
(\lambda,t) e^{i\lambda z} e^{im\varphi} J_m
(\frac{x_{mn}\rho}{a}), \label{a-24}\\
&&\pi(\rho,\varphi,z,t)=\sum_{mn} \int d \lambda  B_{mn} (\lambda,t)
e^{-i\lambda z} e^{-im\varphi} J_m
(\frac{x_{mn}\rho}{a}).\label{aa-24}
\end{eqnarray}
Inserting the above expansions in the symplectic two-form of Eq.
(\ref{w-14}) and using orthogonality of the functions involved,
gives
\be \Omega=2\sum_{mn}2 \pi^2 a^2 K^2_{mn}  d B_{mn}(\lambda,t)
\wedge d A_{mn}(\lambda,t), \label{b-3} \ee
where $K_{mn}=J_{m+1}(x_{mn})$. The non-vanishing brackets among
physical modes can be found by inverting the symplectic matrix of
Eq. (\ref{b-3}) as
\be \left\{A_{mn}(\lambda,t),B_{m'n'}(\lambda',t)\right\}=
\frac{1}{2\pi^2 a^2 K_{mn}^2}\delta_{mm'}
\delta_{nn'}\delta(\lambda-\lambda'). \label{a-26} \ee
Using the above brackets and the expansions (\ref{a-24}) and
\eqref{aa-24} the equal time brackets of the original fields is as
follows;
\begin{eqnarray}&& \left\{\phi (\rho,\varphi,z,t),\pi
(\rho',\varphi',z',t)\right\}=\hspace{8cm} \nonumber \\ &&
\hspace{30mm} \sum_{mn} \frac{1}{2\pi^2  a^2 K_{mn}^2}\int d\lambda
e^{i \lambda(z-z')} e^{im(\varphi-\varphi')} J_m \frac{(x_{mn}
\rho)}{a} J_m \frac{(x_{mn} \rho')}{a} \hspace{-2cm} \label{z-1}
\end{eqnarray}
which is in fact equivalent to a delta function inside the tube and
zero on the boundary, similar to Eq.\eqref{bb-19}.

\section{Klein-Gordon field inside a sphere}
Consider the Klein-Gordon field inside a sphere of radius $R$. The
canonical Hamiltonian is again as in Eq.(\ref{a-4}), while the
primary constraint is given by $ \phi(R,\theta,\varphi)\approx0 $.
Using the total Hamiltonian
\be H_T=H_c+\int d\theta d\varphi\lambda(\theta,\varphi)
\phi(R,\theta ,\varphi)\ , \label{a-30} \ee
the consistency procedure gives, similar to the cylindrical
coordinates, two infinite sets of constraints as
\be (\nabla^2)^n \phi|_{r=R}\approx 0\ ,\hspace{1cm} (\nabla^2)^n
\pi|_{r=R}\approx 0\ . \label{a-31} \ee
According to the spherical symmetry of the problem, we expand our
fields inside the sphere, in their most general form, as the
following Bessel-Fourier expansions
\begin{eqnarray}
&&\phi(r,\theta,\varphi,t)=\sum_{l=0}^\infty \sum_{m=-l}^l
\int_0^\infty d k A_{lm} (k,t)
j_l (k r)  Y_{lm} (\theta,\varphi), \label{aa-32} \\
&&\pi(r,\theta,\varphi,t)=\sum_{l=0}^\infty \sum_{m=-l}^l
\int_0^\infty d k  B_{lm} (k,t) j_l (k r)  Y_{lm}(\theta,\varphi).
\end{eqnarray} \label{a-32}
where $j_l(x)$'s are spherical Bessel functions and $Y_{lm}$'s are
spherical harmonics. We can find the reduced phase space of the
system by imposing the set of constraints (\ref{a-31}) on the
expansion of the fields. Using the spherical Bessel equation we
find, for $\phi$ in Eq.(\ref{aa-32}),
\be \sum_{lm}\int dk A_{lm}(k,t)(k^2 )^j \left[j_l(kr)\right]_{r=R}
Y_{lm}(\theta ,\varphi)=0\ . \ee
This equation for arbitrary $j$ shows that $A_{lm}(k,t)$ should
vanish expect for
\be j_l (k r)|_{r=R}=0 \Rightarrow k_{ln}\equiv \frac{x_{ln}}{R}
\ee
where $x_{ln}$ for positive integers $n$ are roots of the
spherical Bessel function $j_l(x)$. The same thing happen for
$\pi(r,\theta,\varphi,t)$, and we can write the following
expansion for fields in the reduced phase space
\begin{eqnarray}
\phi(r,\theta,\varphi,t)&=&\sum_{lmn}  A_{lmn} (t)
j_l \left(x_{ln}r/R\right)  Y_{lm}(\theta,\varphi) \ ,\label{a-34} \\
\pi(r,\theta,\varphi,t)&=&\sum_{lmn} B_{lmn} (t) j_l \left(x_{ln}
r/R\right)  Y_{lm}(\theta,\varphi) \ .\label{aa-34}
\end{eqnarray}
Again we quantize the theory by using the symplectic approach. Using
the orthogonality of spherical harmonics and spherical Bessel
functions, the symplectic two-form (\ref{w-14}) turns out to be
\be \Omega=\sum_{lmn}R^3 [j_{l+1}(x_{ln})]^2dB_{lmn}\wedge dA_{lmn}
. \ee
Inverting the symplectic matrix, the non-vanishing brackets among
physical modes are
\be \left\{A_{lmn},B_{l'm'n'}\right\}= \frac{2}{ R^3[ j_{l+1}
(x_{ln})]^2}\delta_{ll'} \delta_{mm'} \delta_{nn'}.  \ee
Using the above bracket and Eqs.\eqref{a-34} and (\ref{aa-34}), one
can show the Dirac brackets of the original coordinate and momentum
field are as follows
\be \left\{\phi (r,\theta,\varphi,t),\pi (r',\theta',
\varphi',t)\right\}=\sum_{lmn} \frac{2}{ R^3 [j_{l+1}(x_{ln})]^2}
j_l \frac{(x_{ln} r)}{R} j_l\frac{(x_{ln} r')}{R}
Y_{lm}(\theta,\varphi)Y_{lm}(\theta',\varphi') , \ee
which is equal to delta function inside the sphere and vanishes on
the boundary similar to Eq. (\ref{bb-19}).

\section{Electromagnetic field in a rectangular box}
In the subsequent sections we find the reduced phase space of the
Electromagnetic field in a finite volume. We begin with a
rectangular box, however, before that we consider some general
aspects of the electromagnetic field in Hamiltonian formalism.

The classical Lagrangian of the free electromagnetic field is given
as
\be L=-\frac{1}{4}\int F_{\mu \nu}F^{\mu \nu} d^3 x\ , \label{c-1}
\ee
where $F_{\mu \nu}=\partial_\mu A_\nu-\partial_\nu A_\mu$. The
canonical momenta are $ \pi_\mu (x)=\partial L/\partial {\dot A}^\mu
(x)=-{F^0}_\mu $, giving $\phi_1 \equiv \pi_0$ as the primary
constraint. The basic Poisson bracket among the canonical variables
are
\begin{eqnarray}
&&\{A^\mu(x),\pi_\nu(x')\}=\delta^{\mu}_{\nu} \delta^3(x-x'), \nonumber \\
&&\{A^\mu(x),A^\nu(x')\}= 0, \label{c-4}\\
&&\{\pi_\mu(x),\pi_\nu(x')\}= 0 .\nonumber
\end{eqnarray}
The canonical Hamiltonian reads
\be H_c=\int (\frac{1}{2}\pi_i \pi_i+\pi_i \partial_i
A_0+\frac{1}{4} F_{ij}F_{ij}) d^3 \bx \ .\label{c-6} \ee
The total Hamiltonian is $ H_T=H_c+\int d^3\bx u(\bx,t)
\phi_1(\bx,t), $ where $u(\bx,t)$  is the Lagrange multiplier.
Consistency condition of primary constraint gives the secondary
constraint $\phi_2 \equiv \left\{\phi_1,H_T\right\}= -\partial_i
\pi_i $. The constraints $\phi_1$ and $\phi_2$ are first class, i.e.
$\{\phi_1,\phi_2\}=0$. Consistency of the constraint $\phi_2$ is
satisfied identically, hence no more constraint would emerge.

We fix the gauge generated by $\phi_2$ by imposing the gauge fixing
conditions $ \Omega_2\equiv\partial_i A_i\approx 0$. Consistency of
the gauge fixing condition $ \Omega_2$ then gives the next gauge
fixing condition as $\Omega_1\equiv A_0\approx 0$ which in turn
fixes the gauge generated by the constraint $\phi_1$. We denote the
four constraints $\phi_1$, $\phi_2$, $\Omega_1$ and $\Omega_2$,
which emerge due to the singular structure of the Lagrangian and are
valid throughout the volume of the system, as the "bulk
constraints". Besides the bulk constraints there are also the
"boundary constraints" emerging from the geometry and physical
properties of the boundaries.

Now consider the electromagnetic field inside a rectangular cubic
box with boundaries at $x=0,a,\ y=0,b,\ z=0,c .$ Suppose the walls
are ideal conductors with infinite magnetic permeability, so the
constraints due to boundary conditions are as follows
\begin{eqnarray}&&A_y(x,y,z)|_{x=0,x=a}=A_z(x,y,z)|_{x=0,x=a}=0,
\hspace{1cm} \frac{\partial A_z}{\partial y} -\frac{\partial
A_y}{\partial z}|_{x=0,x=a}=0\ , \nonumber\\
&&A_x(x,y,z)|_{y=0,y=b}=A_z(x,y,z)|_{y=0,y=b}=0,\hspace{1cm}
\frac{\partial A_z}{\partial x}-\frac{\partial A_x}{\partial
z}|_{y=0,y=b}=0 \ ,\label{c-9}\\ &&
A_x(x,y,z)|_{z=0,z=c}=A_y(x,y,z)|_{z=0,z=c}=0,\hspace{1cm}
\frac{\partial A_x}{\partial y}-\frac{\partial A_y}{\partial
x}|_{z=0,z=c}=0 \ .\nonumber
\end{eqnarray}
We denote the above constraints as $\psi_i \hspace{3mm} i=1,\dots
18$. We should then investigate the consistency of the boundary
constraints, together with applying the bulk constraints.
Mathematically we are free to do this in any order we desire. We use
this possibility to follow the simplest way to investigate the
consistency of the constraints and avoid a large amount of useless
algebra. For this reason we omit first the fields $\pi_0$ and $A_0$
from the very beginning to simplify the problem. Hence, the new
total Hamiltonian is
 \be
H_T=\int d^3 \bx \left(\frac{1}{2}\pi_i \pi_i+\frac{1}{4}
F_{ij}F_{ij}\right) + \sum_{i=1}^{18}
\int_{\textrm{boundaries}}\lambda_i \psi_i.\label{c-10}
 \ee
In the last term of the Eq.(\ref{c-10}) the $i=1$ term, for
instance,  is understood as $\int d y d z \lambda_1(y,z)A_y(0,y,z)$
and so on. Using the Poisson brackets (\ref{c-4}), consistency of
the constraint $\psi_1(y,z)=A_y(0,y,z)$ gives the new constraint $
\chi_1 (y,z) \equiv\{A_y(0,y,z), H_T\}=\pi_y(0,y,z)\approx 0$. Hence
we find eighteen second level constraints $\chi_i$ similar to
primary constraints of Eqs.(\ref{c-9}) for momentum fields
$\pi(x,y,z)$. Consistency of $\pi_y(0,y,z)$ then gives $(\partial_y
\partial_i) A_i-\nabla^2A_x$ (while determining the corresponding
Lagrange multiplier $\lambda_1$ to be zero, as discussed for the
Klein-Gordon field after the relation (\ref{a-10})). Since
$\partial_iA_i$ is the bulk constraint $\Omega_2$, consistency of
$\chi_1$ gives $\nabla^2 \psi_1$ at the same boundary.

The consistency procedure leads in this way to infinite chains of
constraints at the boundaries which are obtained by acting
$(\nabla^2)^n$ on the constraints $\psi_i$ in (\ref{c-9}) as well as
similar constraints $\chi_i$ for momentum fields.

In order to impose the constraints we use the usual Fourier
transformations of the real fields as combinations of sine and
cosine  terms. Imposing the boundary constraints
$(\nabla^2)^n\psi_i$ for arbitrary $n$ as well as the bulk
constraint $\partial_iA_i$ leads to remaining special products of
sine and cosine terms as the following combinations of discrete
modes
\begin{eqnarray}
&&A_x(\bx,t)=\sum_{\textbf{k}}a_x(\textbf{k},t)\cos(k_xx)
\sin(k_yy)\sin(k_zz)\ ,\nonumber\\
&&A_y(\bx,t)=\sum_{\textbf{k}} a_y(\textbf{k},t)\sin(k_xx)
\cos(k_yy)\sin(k_zz)\ , \label{cc-15} \\
&&A_z(\bx,t)=\sum_{\textbf{k}} a_z(\textbf{k},t)\sin(k_xx)
\sin(k_yy)\cos(k_zz)\ ,\nonumber
 \end{eqnarray}
where $\textbf{k}=(m\pi/a,n\pi/b,l\pi/c)$ for integers $m$, $n$, and
$l$ and $\textbf{a.k}=0$ due to the bulk constraints $\partial_iA_i
=0$. Defining the orthonormal set of basis vectors $\hat\epsilon_1
(\textbf{k})$, $\hat\epsilon_2(\textbf{k})$ and $\hat{\textbf{k}}$,
the components of the vector potential $A$ can be written in terms
of the physical modes $a^{1}_{mnl}(t)$ and $a^2_{mnl}(t)$ as follows
 \begin{eqnarray}
&&A_x(\bx,t)=\sum_{mnl}\sum_{i=1}^2
a^i_{mnl}(t)\epsilon_{ix}(\textbf{k}) \cos\frac{m\pi
x}{a}\sin\frac{n\pi y}{b} \sin\frac{l\pi z}{c}\ ,\nonumber\\
&&A_y(\bx,t)=\sum_{mnl} \sum_{i=1}^2
a^i_{mnl}(t)\epsilon_{iy}(\textbf{k})\sin\frac{m\pi
x}{a}\cos\frac{n\pi y}{b} \sin\frac{l\pi z}{c}\ , \label{c-15}
\\&&A_z(\bx,t)=\sum_{mnl} \sum_{i=1}^2 a^i_{mnl}(t)\epsilon_{iz}(\textbf{k})
\sin\frac{m\pi x}{a} \sin\frac{n\pi
y}{b}\cos\frac{l\pi z}{c}\ .\nonumber
 \end{eqnarray}
The same story should be repeated for the momentum fields to get
\begin{eqnarray}
&&\pi_x(\bx,t)=\sum_{mnl}\sum_{i=1}^2
b^i_{mnl}(t)\epsilon_{ix}(\textbf{k}) \cos\frac{m\pi
x}{a}\sin\frac{n\pi y}{b} \sin\frac{l\pi z}{c}\ ,\nonumber\\
&&\pi_y(\bx,t)=\sum_{mnl} \sum_{i=1}^2
b^i_{mnl}(t)\epsilon_{iy}(\textbf{k})\sin\frac{m\pi
x}{a}\cos\frac{n\pi y}{b} \sin\frac{l\pi z}{c}\ ,\label{c-16}
\\&&\pi_z(\bx,t)=\sum_{mnl} \sum_{i=1}^2 b^i_{mnl}(t)\epsilon_{iz}(\textbf{k})
\sin\frac{m\pi x}{a} \sin\frac{n\pi y}{b}\cos\frac{l\pi z}{c}\
.\nonumber
 \end{eqnarray}
Now we can quantize the system be calculating the symplectic
two-form of the system. Using the general formula (\ref{w-14}) and
orthogonality of physical modes given in Eqs.(\ref{cc-15}) we find
\be \Omega =\sum_{i=1}^2\sum_{mnl} \frac{abc}{8}d
b^{i}_{mnl}(t)\wedge d a^{i}_{mnl}(t) \ .\ee
Hence, non-vanishing bracket among the physical modes are as
follows
\be \left[a^{i}_{mnl}(t) , b^{j}_{m'n'l'}(t)\right]=\frac{16}{abc}
\delta^{ij} \delta_{mm'} \delta_{nn'} \delta_{ll'} \ .\label{r-2}
\ee
As is seen the reduced phase space is described by the canonical
conjugate pairs $(a^i_{mnl} , b^i_{mnl})$. Once again we can compute
the Dirac brackets of the original fields as follows
\begin{eqnarray}
&&\left[A^x(\textbf{x},t),\pi_x(\textbf{x}',t)\right]=
\sum_{mnl}\frac{32}{abc} \cos\frac{m\pi x}{a} \cos\frac{m\pi
x'}{a}\sin\frac{n\pi y}{b}\sin\frac{n\pi
y'}{b}\sin\frac{l\pi z}{c}\sin\frac{l\pi z'}{c}\ ,\nonumber\\
&&\left[A^y(\textbf{y},t),\pi_y(\textbf{y}',t)\right]=
\sum_{mnl}\frac{32}{abc} \sin\frac{m\pi x}{a}\sin\frac{m\pi x'}{a}
\cos\frac{n\pi y}{b}\cos\frac{n\pi y'}
{b}\sin\frac{l\pi z}{c}\sin\frac{l\pi z'}{c}\ ,\label{h-3}\\
&&\left[A^z(\textbf{z},t),\pi_z(\textbf{z}',t)\right]=
\sum_{mnl}\frac{32}{abc} \sin\frac{m\pi x}{a}\sin\frac{m\pi x'}{a}
\sin\frac{n\pi y}{b}\sin\frac{n\pi y'}{b}\cos\frac{l\pi
z}{c}\cos\frac{l\pi z'}{c}\ ,\nonumber\end{eqnarray}
which differs from the primary Poisson brackets given in
Eqs.\eqref{c-4} in the sense that the constraints (\ref{c-9}) (and
their partners for the momentum fields) are satisfied strongly on
the corresponding walls. For example $A_y(0,y,z)$ as well as
$[\frac{\partial A_z}{\partial y}-\frac{\partial A_y}{\partial
z}](0,y,z)$ have vanishing Dirac bracket with every thing. Moreover
$\nabla . A$ has vanishing Dirac bracket with all of the fields
everywhere inside the box.

The canonical Hamiltonian of the system can be written in terms of
the physical modes as
\be
H_c=\frac{abc}{16}\sum_{mnl}\sum_{i=1}^2\left[({b_i^{mnl}})^2+\omega
_{mnl}^2 ({a_{1}^{mnl}})^2 \right], \ee
where
\be \omega _{mnl}^2=\frac{m^2 \pi^2}{a^2}+\frac{n^2
\pi^2}{b^2}+\frac{l^2 \pi^2}{c^2}. \ee
In this way the problem in the reduced phase space reduces to a
summation over simple harmonic oscillators obeying the following
dynamics
\begin{eqnarray} && a_i^{mnl}(t)=a_i^{mnl}(0)\cos (\omega_{mnl}
t) +\frac{b_i^{mnl}(0)}{\omega_{mnl}}\sin (\omega_{mnl} t)\ , \label{c-19} \\
&& b_i^{mnl}(t)=-a_i^{mnl}(0)\omega_{mnl}\sin (\omega_{mnl} t) +
b_i^{mnl}(0) \cos (\omega_{mnl} t) \ .\label{cc-19}\end{eqnarray}

Inserting the time dependent modes of \eqref{c-19} and \eqref{cc-19}
into the expansions (\ref{c-15}) gives the vector potential
components in terms of Schrodinger modes $a_i^{mnl}(0)$ and
$b_i^{mnl}(0))$ which obey the same algebra as given in
Eq.\eqref{r-2}. Using $B=\nabla \times A$ and $E=-\partial
A/\partial t$ in the Coulomb gauge considered above, we can write
the physical fields $E(\bx,t)$ and $B(\bx,t)$  as well as every
physical quantity of our interest in terms of Schrodinger modes.
These are also the basic quantum objects upon quantization. Quantum
commutators of the Schrodinger modes come from the fundamental Dirac
prescription as
\be \left[a^{i}_{mnl}(0) , b^{j}_{m'n'l'}(0)\right]=i\hbar
\frac{16}{abc} \delta^{ij} \delta_{mm'} \delta_{nn'} \delta_{ll'} \
.\label{r-2} \ee
\section{ Electromagnetic field in a cylindrical box}

Consider the Electromagnetic field inside a cylindrical box of
radius $R$, and length $d$ made of an ideal conductor. The canonical
Hamiltonian is given by Eq. (\ref{c-6}). The boundary conditions due
to vanishing of the tangental components of electric field are
\be A^{\rho}|_{z=0}=A^{\rho}|_{z=d}=0,\hspace{2mm}
A^{\phi}|_{z=0}=A^{\phi}|_{z=d}=0,\hspace{2mm}
A^{z}|_{\rho=R}=A^{\phi}|_{\rho=R}=0, \label{dd-1} \ee
while vanishing of normal component of the magnetic field gives
\be \frac{1}{\rho}(\frac{\partial A^z}{\partial \phi}-\frac{\partial
A^\phi}{\partial z})|_{\rho=R}=0,\hspace{2mm}
\frac{1}{\rho}(\frac{\partial}{\partial \rho}(\rho
A^\phi)-\frac{\partial A^\rho}{\partial \varphi})|_{z=0,z=d}=0.
\label{d-1} \ee
The fundamental Poisson brackets among field components  can be
written in cylindrical coordinates as
 \be \left\{A^i(\rho ,
 \varphi,z),\pi_j(\rho',\varphi',z'\right\}= \frac{1}{\rho }\delta^i_j\delta
 (\rho -\rho ')\delta (\varphi-\varphi')\delta(z-z'). \label{o-2}
 \ee
Imposing first the bulk constraints $A^0\approx 0$ and $\pi_0
\approx 0$, the total Hamiltonian read
\be H_T=H_c+\sum_{i=1}^{9} \lambda_i \psi_i^0 .\label{o-3} \ee
where the summation is over the nine constraints $\psi_i^0$ given in
Eqs.\eqref{dd-1} and \eqref{d-1}. Consistency of the constraints
$\psi_i^0$ gives a copy of them in terms of momentum fields which we
call them $\chi_i^0$. Consistency of the new constraints should be
achieved by writing the Hamiltonian in terms of cylindrical
coordinate and using the Poisson bracket (\ref{o-2}). The procedure
is as before and straightforward; however, care is needed to
considering the spacial derivatives of unit vectors. Fortunately,
the additional terms are proportional to the previous constraints
which vanish weekly. Similar to previous cases we find two sets of
infinite constraints at the boundaries as
\be \psi _i^n=\nabla^{2n} \psi_i^0\approx 0 ,\hspace{4mm}
\chi_i^{n}=\nabla^{2n} \chi_i^0\approx 0 \label{o-4} \ee
Noting the cylindrical geometry of the problem, we expand our
fields in the most general form as
\begin{eqnarray}
A^\alpha(\rho,\varphi,z,t)& = &\sum_m \int d \lambda d \gamma
A_{m}^\alpha (\lambda,\gamma,t)
e^{i \lambda z} e^{im\varphi} J_m (\gamma \rho) \\
\pi_\alpha (\rho,\varphi,z,t)&= &\sum_m \int d \lambda d \gamma
B_{m}^\alpha (\lambda,\gamma,t) e^{-i \lambda z} e^{-im\varphi} J_m
(\gamma \rho) \label{o-5}
\end{eqnarray}
where $\alpha$ runs over  $\rho,\phi,z.$ and $m$ is an integer,
since the fields should be single valued at each point of space.
Imposing the constraints $\nabla^{2n} A^\alpha |_{\rho=R}$ for
$\alpha=\varphi$ and $z$ gives
\be \sum_m \int d\lambda d\gamma A_{m}^\alpha (\lambda,\gamma,t)
e^{i \lambda z} e^{im \varphi}
\left[\left(-\gamma^2-\lambda^2\right)^n J_m (\gamma
\rho)\right]_{\rho=R} =0  \label{o-6} \ee
for arbitrary $n$. Hence, $A_m^\varphi$ and $A_m^z$ should vanish
expect for $\gamma_{mn}R=x_{mn}$ where $x_{mn}$ is the $n$th root of
the Bessel function $J_m(x)$. Imposing the constraints $\nabla^{2n}
A^\rho|_{z=0,d}=0$ and $\nabla^{2n} A^\varphi|_{z=0,d}=0$  gives the
$z$ dependence of the corresponding fields as $\sin (lz\pi/d) $ for
integer $l$. We should also impose the gauge fixing condition
\be \partial_i A_i\equiv \frac{1}{\rho}\frac{\partial}{\partial
\rho}(\rho A^\rho)+\frac{1}{\rho} \frac{\partial A^\varphi}{\partial
\varphi}+\frac{\partial A^z}{\partial z}=0 \ .\label{o-10} \ee
This can be satisfied at any arbitrary point in space only if the
$\gamma$ and $\lambda$  acquire the same quantized values in all of
the expansions. From Eq.\eqref{o-10} it is clear that the
$z$-dependence of $A^z$ should be of the form $\cos (l\pi z/d)$ and
$\rho$-dependence of $A^\rho$ should be combinations of $J_{m} (
\gamma_{mn} \rho )$ for suitable indices $m$. Functions $J'_{m}(x)$
and $J_{m}(x)/x$ for $x=\gamma_{mn} \rho $ can be written in terms
of $J_{m\pm 1}(x)$ and may turn out to be useful.

Hence, imposing all of the requirements of the constraints
\eqref{o-4} and \eqref{o-10} leads to determining $A^\rho$,
$A^\varphi$ and $A^z$ in terms of TM modes,
\begin{eqnarray}
Q^\rho_{mnl}(\rho,\varphi,z,t)&=&- \frac{l\pi}{d\gamma_{mn}} \sin
\frac{l\pi z}{d} e^{im\varphi} J'_{m} (\gamma_{mn}
\rho),\nonumber\\
Q^\varphi_{mnl}(\rho,\varphi,z,t)&=&
-\frac{l\pi}{d\gamma^{2}_{m,n}}\frac{im}{\rho} \sin \frac{l\pi z}{d}
e^{im\varphi} J_{m}(\gamma_{mn} \rho), \label{o-9} \\
Q^z_{mnl}(\rho,\varphi,z,t)&=& \cos \frac{l\pi z}{d} e^{im\varphi}
J_{m}(\gamma_{mn} \rho ), \nonumber
\end{eqnarray}
and TE modes,
\begin{eqnarray}
R^\rho_{mnl}(\rho,\varphi,z,t)&=& -\frac{im}{\gamma_{mn}\rho} \sin
\frac{l\pi z}{d} e^{im\varphi} J_{m} (\gamma_{mn}
\rho),\nonumber\\
R^\varphi_{mnl}(\rho,\varphi,z,t)&=& \sin \frac{l\pi z}{d}
e^{im\varphi} J'_{m}(\gamma_{mn}
\rho),\label{oo-9} \\
R^z_{mnl}(\rho,\varphi,z,t)&=&0. \nonumber
\end{eqnarray}
The general form of dynamical fields $\textbf{A}(\rho,\varphi,z,t)$
in terms of physical modes is
\be \textbf{A}(\rho,\varphi,z,t)=\sum_{m ,n,l}  A_{mnl}^{1}(t)
\textbf{Q}_{mnl}(\rho,\varphi,z)+ A_{mnl}^{2}(t)
\textbf{R}_{mnl}(\rho,\varphi,z). \label{sh1} \ee
Similar results should be considered for the momentum field $\pi$ as
\be \pi(\rho,\varphi,z,t)=\sum_{m ,n,l}  B_{mnl}^{1}(t)
\textbf{Q}^\ast_{mnl}(\rho,\varphi,z)+ B_{mnl}^{2}(t)
\textbf{R}^\ast_{mnl}(\rho,\varphi,z) \label{sh2} \ee
Hence our reduced phase space is described by variables
$A_{mnl}^{i}(t)$ and $B^i_{mnl}$ for $i=1,2$. Using the equations
$E=-\partial A/\partial t$ and $B=\nabla \times A$ in the coulomb
gauge (which is used here) one can show that the above modes are
consistent with the standard results given in the text books.

Now we can quantize the system by using the symplectic two-form
given in formula (\ref{w-14}). Fortunately the physical modes
$\textbf{Q}_{mnl}(\rho,\varphi,z)$ and
$\textbf{R}_{mnl}(\rho,\varphi,z)$ constitute an orthogonal set of
vector functions inside the cylindrical box. Using orthogonality
conditions
\begin{eqnarray} \int d^3\textbf{x}
\textbf{Q}^\ast_{mnl}(\textbf{x}).\textbf{Q}_{m'n'l'} (\textbf{x})
&=& \frac{V}{2} K_{mn}^2 \frac{\omega_{mnl}^2} {\gamma_{mn}^2}
\delta_{mm'}\delta_{nn'}\delta_{ll'} \label{sh13}\\ \int
d^3\textbf{x} \textbf{R}^\ast_{mnl}(\textbf{x}).\textbf{R}_{m'n'l'}
(\textbf{x}) &=& \frac{V}{2} K_{mn}^2
\delta_{mm'}\delta_{nn'}\delta_{ll'} \label{sh14}
\end{eqnarray}
we find
\be \Omega = \sum_{mnl} V K_{mn}^2 \left[
\frac{\omega_{mnl}^2}{\gamma_{mn}^2} d B_{mnl}^{1} \wedge d
A_{mnl}^{1} + d B_{mnl}^{2} \wedge d A_{m nl}^{2} \label{sh3}
\right]\ , \ee
where $\omega_{mnl}^2 = \gamma_{mn}^2+ l^2\pi^2/d^2$,
$K_{mn}=J_{m+1}( x_{mn})$ and $V$ is the volume of the cylinder.
Since the symplectic matrix is diagonal in the basis of TM and TE
modes, it is an easy task to invert it and read the non-vanishing
Dirac brackets among physical modes as follows
\begin{eqnarray}
 \left[A^1_{mnl}(t),B^1_{m'n'l'}(t)\right]&=& \frac{2}{VK_{mn}^2}
\frac{\gamma_{mn}^2}{\omega_{mnl}^2}\delta_{mm'}\delta_{nn'}
\delta_{ll'}\nonumber\\
\left[A^2_{mnl}(t),B^2_{m'n'l'}(t)\right]&=&\frac{2}{VK_{mn}^2}
\delta_{mm'}\delta_{nn'} \label{sh15} \delta_{ll'}
\end{eqnarray}
Note once again that we have not solved the equations of motion
completely; so that the time dependence of the physical modes
$A^i_{nml}(t)$ and $B^j_{nml}(t)$ are not determined yet. This,
however, can be achieved by writing the canonical Hamiltonian in
terms of physical modes as follows
\be H = \sum_{mnl} \frac{V}{2}K^2_{mn}\left[
\frac{\omega_{mnl}^2}{\gamma_{mn}^2} \left( \omega_{mnl}^2
{A^1_{mnl}}^2+ {B^1_{mnl}}^2 \right) +\left(\omega_{mnl}^2
{A^2_{mnl}}^2+ {B^2_{mnl}}^2 \right) \right] \ee
Solving the equations of motion for physical modes gives, as in the
previous cases
\begin{eqnarray}
\dot{A^i} &=&  [A^i_{mnl},H]=B^i_{mnl}\nonumber \\
\dot{B^i} &=& [B^i_{mnl},H]= -\omega^2_{mnl} A^i_{mnl}
\end{eqnarray}
for $i=1,2$. By straightforward calculation using the Dirac brackets
of Eqs. (\ref{sh15}) and the following completeness relations (for
divergence-less vector functions satisfying our boundary conditions)
of physical modes
\be \sum_{mnl} \frac{2}{VK_{mn}^2} \left[
\frac{\gamma_{mn}^2}{\omega_{mnl}^2} \textbf{Q}^{\ast
\alpha}_{mnl}(\textbf{x}) \textbf{Q}_{mnl}^\beta (\textbf{x'}) +
\textbf{R}^{\ast \alpha}_{mnl}(\textbf{x}) \textbf{R}_{mnl}^\beta
(\textbf{x'}) \right] = \delta^{\alpha \beta}\delta(\textbf{x-x'})
\label{sh16} \ee
where $\alpha$ and $\beta$ refer to different components in
cylindrical coordinates, we can find the Dirac brackets of the
fields \textbf{inside the cylindrical box} as
\be \left\{A^\alpha (\bx,t),\pi^\beta (\bx',t)\right\}=
\delta^{\alpha \beta}\delta(\textbf{x-x'}). \label{b-19} \ee
However, the result is not the same as the Poisson brackets
(\ref{o-2}) which we begin with. In fact special construction
(\ref{sh1}) and (\ref{sh2}) of fields in terms of physical modes
(\ref{o-9}) and (\ref{oo-9}) shows that the Dirac brackets on the
walls are consistent with the constraints (\ref{dd-1}) and
(\ref{d-1}). For example $\left\{ A^\rho (\bx,t),\pi^\beta
(\bx',t)\right\}$, $\left\{A^\phi (\bx,t),\pi^\beta
(\bx',t)\right\}$ and
$\left\{\frac{1}{\rho}(\frac{\partial}{\partial \rho}(\rho
A^\phi)-\frac{\partial A^\rho}{\partial \varphi}) (\bx,t),\pi^\beta
(\bx',t)\right\}$ vanish on the end faces $z=0$ and $z=d$ while
$\left\{\nabla .\textbf{A} (\bx,t),\pi^\beta (\bx',t)\right\}$
vanishes everywhere inside the box and on the walls.

\section{Electromagnetic field inside a sphere}
Consider the Electromagnetic field inside a sphere of radius $R.$
Assuming the interior wall of the sphere is an ideal conductor, the
boundary conditions are
\be A^{\phi}|_{r=R}=0,\hspace{2mm}
A^{\theta}|_{r=R}=0,\hspace{2mm} \frac{1}{r\sin
\theta}[(\frac{\partial}{\partial \theta}(\sin \theta A
^{\varphi})-\frac{\partial A^\theta}{\partial \varphi}]|_{r=R}=0
\label{e-2} \ee
The total Hamiltonian in the gauge $A^0=0$ is $ H_T =H_c+
\sum_{i=1}^{3} \lambda_i \psi^0_i$, where the summation is over the
three boundary constraints given by Eq. \eqref{e-2}. The fundamental
Poisson brackets among the field components read
\be \left\{A^\alpha(r ,\theta, \varphi),\pi_\beta
(\rho',\theta',\varphi')\right\}= \frac{\delta^\alpha_\beta}{r^2
\sin^ 2 \theta}\delta (r -r ')\delta(\theta-\theta')\delta
(\varphi-\varphi'), \label{e-3} \ee
where $\alpha$ and $\beta$ run over $r$, $\theta$ and $\varphi$. The
procedure of consistency of the constraints leads to two infinite
sets of constraints as follows
\be \nabla^{2n} \psi^0_i=0,\nabla^{2n}\chi^0_i=0 \label{e-4} \ee
where $\chi^0_i$ are similar to $\psi^0_i$ of Eq.\eqref{e-2} in
terms of momentum fields $\pi_\alpha$. Let impose the constraints on
the most general form of Bessel-Fourier expansions as
\be A^\alpha (r,\theta,\varphi,t)=\sum_{lm} \int dk A_{lm}^\alpha
(k,t) j_l (kr)  Y_{lm}(\theta,\varphi) \ . \ee
Using the differential equation of spherical Bessel functions
as
\be \left(\frac{d^2}{d r^2}+\frac{2}{r}\frac{d}{d
r}-\frac{l(l+1)}{r^2}+k^2 \right) j_l (k r)=0, \ee
the constraints $\nabla^{2n}A^\theta=0$ and
$\nabla^{2n}A^\varphi=0$ at $r=R$ give
\be \sum_{lm} \int d k A_{lm}^\alpha(k,t)\left[(-k^2)^n j_l (k
r)\right] _{r=R}Y_{lm}(\theta,\varphi) =0. \label{e-5} \ee
Hence $A_{lm}(k,t)$ should vanish expect for
\be j_l (kr)|_{r=R}=0 \Rightarrow k_{ln}\equiv \frac{x_{ln}}{R}
\ee
where $x_{ln}$ are roots of the equation $j_l(x)=0$. Up to this
stage,  we can write the following expansion for the first two
components of the $\textbf{A}$-field
\begin{eqnarray}
A^\theta(r,\theta,\varphi)&=&\sum_{lmn}  A^{\theta}_ {lmn}(t)
j_{l} (k_{ln} r) Y_{lm}(\theta, \varphi)\nonumber\\
A^\varphi(r,\theta,\varphi)&=&\sum_{lmn}  A^{\varphi}_{lmn}(t)
j_{l} (k_{ln} r) Y_{lm}(\theta, \varphi)
\end{eqnarray}
The gauge fixing condition $\nabla .\textbf{A}=0$ in spherical
coordinate, i.e.
\be \frac{1}{r^2}\frac{\partial}{\partial r}(r^2
A^r)+ \frac{1}{r \sin \theta}\frac{\partial}{\partial \theta}(\sin
\theta A^\theta) +\frac{1}{r \sin \theta}\frac{\partial
A^\varphi}{\partial \varphi}=0 \label{e-6} \ee
can be satisfied at any point, if the variable $k$ in the expansion
of $A^r$ is of the form $x_{ln}/R$, i.e.
\be   A^r (r,\theta,\varphi)= \sum_{lmn}  A^{r}_{lmn} (t) j_{l}
(k_{ln} r) Y_{lm}(\theta,\varphi) \ee
Hence, the final answer for each $l$ is the product of
$j_l(x_{ln}r/R)$ with some suitable vector combination of $Y_{lm}$'s
which satisfy the gauge $\nabla .\textbf{A}=0$. This leads us
naturally to the notion of "vector spherical harmonics". As is
well-known from the standard text books on special functions, for
each $l,m$ there exist three orthogonal vector combinations
$X_{lm}$, $U_{lm}$ and $V_{lm}$, which are vector eigenfunctions of
the angular part of Laplacian  operator; among them only $X_{lm}$
is the suitable one such that the combination
$\textbf{A}_{lm}(r,\theta,\varphi)\equiv \sum_m \textbf{a}_{lm}
j_l(kr)Y_{lm} (\theta,\varphi)$ satisfies $\nabla
.\textbf{A}_{lm}=0$. This function is defined as
\be X_{lm}=\hat{\theta} \left\{\frac{-m}{[l(l+1)]^
\frac{1}{2}\sin\theta}Y_{lm}\right\}
+\hat{\varphi}\left\{\frac{-i}{[l(l+1)]^\frac{1}{2}} \frac{\partial
Y_{lm}}{\partial \theta}\right\}. \ee

This is the only reasonable answer. In other words, no other mode
can be found that satisfies all of our physical conditions. This is
specular for spherical coordinates which the gauge condition $\nabla
.\textbf{A}=0$ leaves us just with one mode (in each pair $lm$) as
follows
\be \textbf{A}(r,\theta,\varphi)=\sum_{lmn} A_{lmn} j_l(x_{ln}r/R)
X_{lm}(\theta,\varphi). \label{f-3} \ee
Similar results should be written for the momentum fields in terms
of the physical modes $B_{lmn}$. Using the orthogonality relations
\begin{eqnarray} & &  \int X_{lm}(\theta,\varphi).X_{l'm'}(\theta,\varphi)d \Omega =
\delta_{ll'}\delta_{mm'}. \nonumber \\ & & \int j_l(x_{ln}r/R)
j_l(x_{ln'}r/R)r^2 dr=[j_{l+1}(x_{ln }r/R)]^2 \label{f-2}
\end{eqnarray}
and the general formula (\ref{w-14}) for the symplectic two-form, we
find
\be \Omega =R^3j^2_{l+1}(x_{ln}) dB_{mnl}^\theta(t) \wedge d
A_{mnl}^\theta(t) \ee
Fortunately the symplectic two-form is in the diagonal form between
conjugate pairs $(A_{lmn},B_{lmn})$. Hence it can be inverted easily
to give the following brackets among physical modes,
\be \left[A_{mnl}(t),B_{m'n'l'}(t)\right]= \frac{2}{R^3
j^2_{l+1}(x_{ln })}\delta_{ll'} \delta_{mm'}\delta_{nn'}. \ee
Finally for Dirac brackets  of the original fields we have $
\left[A^\alpha (r,\theta,\varphi,t),\pi_\beta
(r',\theta',\varphi',t')\right]= \delta^\alpha_\beta
\delta^3(\bx-\bx')$, inside the box; while the Dirac brackets of the
constraints vanish strongly on the surface of the sphere. Moreover,
$\nabla.A$, $\nabla.\pi$, as well as $A^0$ and $\pi_0$ have
vanishing Dirac brackets with all of the fields.

\section{Conclusions}

Our aim in this paper was presenting a systematic method for
quantization fields with given boundary conditions on the walls of a
box which the fields live in it. We avoid proposing different ad-hoc
quantization assumptions for individual models. Instead, we think
the brilliant prescription of Dirac (i.e. converting the Dirac
brackets to commutators) is the only needed tool for this reason. In
other words, it is not just a criterion to be satisfied by a
quantization assumption; it is, on the other hand, a road-map for
quantizing every desired model.

Traditionally quantization of the fields is achieved by assuming
certain commutation relations among the coefficients of the
expansions of the fields in terms of solutions of the equations of
motion and then showing that the assumed commutation relation gives
the standard canonical commutation relations among the original
fields. We have three objections against this way of thinking.
First, there is not enough logics in this point of view, by its own.
In fact, the main stress is on the standard canonical commutation
relations coming from the classical Poisson brackets. Second, there
is no guaranty about the uniqueness of the method, since the
classical brackets are not the beginning point of the quantization
procedure. Why no other quantization assumption can be proposed
which are consistent with the same set of canonical brackets?

Finally, the third and most important objection is inherent in this
question: "which set of classical brackets should be resulted from
the assumed commutation relations among the coefficients, Poisson
brackets or Dirac brackets?". Text books often refer to Poisson
brackets in the bulk of the medium. However, in the presence of
boundary conditions this is no more correct; since the Poisson
brackets in the bulk are not consistent with the boundary
conditions. For example, when you are given a Dirichlet boundary
condition, the vanishing field can no more give delta function
(unity) in its bracket with the corresponding momentum field. As we
know in the context of constrained systems, in the presence of
second class constraints (which is the case for boundary conditions
as Dirac constraints) the Poisson brackets should be replaced with
Dirac brackets before quantization.

We show in this paper, in almost all of the examples, that the Dirac
brackets are different with the Poisson brackets which we begin with
(see for instance Eqs. \ref{b-18} and \ref{bb-19} and \ref{h-3}).
The consistency procedure of Dirac, together with omitting the modes
which are associated with the second class constraints are essential
steps to obtain classical brackets which are consistent with the
given boundary conditions. The second step is equivalent to
considering the Dirac brackets instead of Poisson brackets. Hence,
the quantization postulates should be based on the Dirac brackets of
the fields. This simple point is not as clear and well-known as it
may seem. In fact concentrating on the contradiction between the
canonical commutators and the mixed boundary conditions in string
theory observed by Seiberg and Witten \cite{witten} leaded to a
stream of papers on non commutativity on the brain coordinates
linking by bosonic strings.

We used here a special approach to Dirac method based on finding the
symplectic matrix in the basis of "physical modes" and inverting it
to find the (Dirac) brackets among them. By physical modes we mean
degrees of freedom which are compatible with the constraints of the
system including first class constraints (as generators of gauge
transformations), gauge fixing conditions, intrinsic second class
constraints, and finally primary boundary conditions as well as
secondary conditions on the boundaries emerging from their
consistency in time. This does not mean "the solutions of the
equations of motion", since the latter depends on the particular
form of the Hamiltonian of the system.

We considered six examples. In each case we were able to find
"suitable basis" for the space of physical variables, so that
imposing the constraints leaded to omitting a number of conjugate
pairs as second class constraints. Hence, we were left with a
reduced phase space with a prescribed canonical basis, which
converted to canonical operators upon quantization. One important
point is that the Fourier modes are not necessarily the canonical
operators of a quantum theory; this is only the case for a geometry
consistent with cartesian coordinates. Note that in the general case
there is no guaranty  to find a "suitable basis". Orthogonality
plays an important role in this regard. If the physical modes are
orthogonal the symplectic matrix, in p-q and q-p blocks, would be
diagonal. Hence, the Dirac brackets coming from inverting the
symplectic two-form (see Eq. \ref{w-16}) will be canonical. This
means that we have been lucky enough to find the canonical basis of
the reduced phase space. However, on the basis of the famous Darboux
theorem \cite{mojiri}, we are only sure about the existence of a
canonical basis in the reduced phase space.

\textbf{Acknowledgment}

The authors would like to thank Hamed Ghaemi and Sara Aghababaei for
reading the manuscript and checking our calculations.

\end{document}